# Power and Packet Rate Control for Vehicular Networks in Multi-Application Scenarios

Miguel Sepulcre, Javier Gozalvez and M. Carmen Lucas-Estañ

*Abstract*—Vehicular networks require vehicles to periodically transmit 1-hop broadcast packets in order to detect other vehicles in their local neighborhood. Many vehicular applications depend on the correct reception of these packets that are transmitted on a common control channel. Vehicles will actually be required to simultaneously execute multiple applications. The transmission of the broadcast packets should hence be configured to satisfy the requirements of all applications while controlling the channel load. This can be challenging when vehicles simultaneously run multiple applications, and each application has different requirements that vary with the vehicular context (e.g. speed and density). In this context, this paper proposes and evaluates different techniques to dynamically adapt the rate and power of 1-hop broadcast packets per vehicle in multi-application scenarios. The proposed techniques are designed to satisfy the requirements of multiple simultaneous applications and reduce the channel load. The evaluation shows that the proposed techniques significantly decrease the channel load, and can better satisfy the requirements of multiple applications compared to existing approaches, in particular the Message Handler specified in the SAE J2735 DSRC Message Set Dictionary.

## I. INTRODUCTION

Vehicular networks or cooperative ITS systems will enable a wide range of active safety and traffic management applications, e.g. intersection collision warning, forward collision warning or lane change assistance to name a few [1]. These applications require vehicles to periodically broadcast and receive packets with positioning and status information using for example IEEE 802.11p/DSRC [2] or ETSI ITS-G5 [3] at the 5.9GHz band. These packets are known as CAM (Cooperative Awareness Messages) in Europe and BSM (Basic Safety Messages) in the US. CAMs or BSMs will be transmitted on the so-called control channel. The control channel is a reference channel to detect the presence of neighbouring vehicles. Such detection is necessary to execute many vehicular applications. IEEE 802.11p utilizes the Carrier Sense Multiple Access Scheme (CSMA) medium access mechanism and is therefore prone to network instability under high traffic loads. Given the reference nature of the control channel, mechanisms are necessary to control the load on the control channel and ensure the network stability. These mechanisms include, for example, the adaptation of the number of packets transmitted per second by each vehicle and their transmission power [4].

Cooperative ITS standards define different communication requirements for each vehicular application [5]. These requirements are specified in terms of metrics such as the communication range, packet transmission/reception rate, reliability or transmission/reception latency [5][6]. Different studies [7]-[10] have shown that these requirements should depend on context factors such as the vehicle's speed, the direction of movement, the driver's reaction time or the position of neighboring vehicles. In addition, vehicles will need to simultaneously run different applications, and each of them could have different requirements. Fig. 1 illustrates an example where a vehicle approaching an intersection needs to simultaneously execute multiple applications due to the presence of different nearby vehicles. The vehicle might be required to transmit 1-hop broadcast packets at high power to satisfy the requirements of the intersection collision warning and left turn assistance applications, and be able to detect potentially colliding vehicles at the intersection (especially under Non-Line-of-Sight conditions). On the other hand, the forward collision warning and lane change assistance applications in the example do not require transmitting 1-hop broadcast packets at high power, but require instead a higher packet rate given the proximity of potentially colliding vehicles. In this case, it is not evident what should be the best configuration of communication parameters, and techniques are necessary to determine under multi-application scenarios how many packets should be transmitted per second, and what should be their power.

Different techniques have been proposed in the literature to dynamically adapt the transmission parameters of 1-hop broadcast packets in order to satisfy the application requirements and control the channel load. However, most proposals do not consider the fact that vehicles will need to simultaneously run multiple applications with different requirements, or do not fully exploit the capacity of IEEE

This work was supported in part by the Spanish Ministry of Economy and Competitiveness and FEDER funds under project TEC2017-88612-R, and by the TransAID project under the Horizon2020 Framework Programme, Grant Agreement no. 723390. Miguel Sepulcre, Javier Gozalvez and M. Carmen Lucas-Estañ are with the UWICORE Laboratory, Universidad Miguel Hernandez de Elche (UMH), Spain. E-mail: msepulcre@umh.es, j.gozalvez@umh.es, m.lucas@umh.es.

802.11p to adapt the packet rate and power on a per-packet basis. This paper proposes and evaluates novel techniques to dynamically configure the transmission parameters of 1-hop broadcast packets (packet rate and power) in multi-application scenarios. The goal is to satisfy the requirements of multiple simultaneous applications, and reduce the channel load in order to improve the network stability.

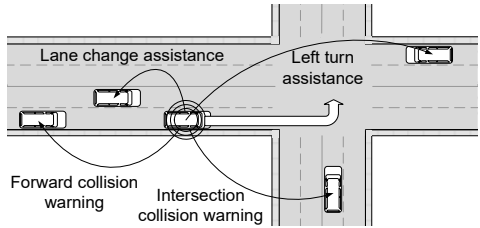

Fig. 1. Example of a multi-application scenario.

## II. RELATED WORK

Various studies [11]-[13] have individually analyzed the performance of different cooperative vehicular applications, and shown that the communication parameters of vehicles should be carefully configured and adapted, in particular under adverse conditions (e.g. high speed or high channel load). Several techniques have been reported in the literature to dynamically adapt the transmission parameters of CAM/BSM packets. For example, [14] proposes a control algorithm that adapts the packet rate of each vehicle to satisfy the demanded tracking accuracy. To this aim, vehicles continuously estimate the potential tracking error of neighboring vehicles, and the transmission of a new CAM/BSM is only triggered when a certain error threshold is surpassed. [15] proposes an algorithm to adapt the communication parameters of each vehicle taking into account their application requirements and the channel load on the control channel. The authors propose in [16] an opportunistic scheme that adapts the rate and power of CAMs/BSMs based on the distance of vehicles to critical safety areas such as intersections. Similarly, the work in [17] proposes to adapt the rate of CAMs/BSMs at intersections based on the intersection's collision probability, which is estimated using received CAMs/BSMs. The work in [18] proposes an algorithm to adapt the CAM/BSM rate and power taking into account application and network level performance metrics. To do so, vehicles share their maximum tolerable time between periodic packets as well as the channel load. The study in [19] formulates the problem of adapting each vehicle's communication parameters as a network utility maximization problem. The utility of a network link depends on the expected packet delay and the vehicles' driving context. [19] defines as driving context the distance between vehicles and the relative speed, and demonstrates that their technique can prioritize transmissions from safety-critical vehicles.

To the authors' knowledge, the only study that considers scenarios where vehicles simultaneously run multiple applications was presented in [20]. It proposes a Message Dispatcher that avoids unnecessarily duplicating BSMs when different safety applications simultaneously run in a vehicle. If different applications require the transmission of the same Data Elements (DE) but with diverse transmission frequencies, the Message Dispatcher generates the lowest number of packets (with minimum possible number of DEs) required to satisfy the packet rate requirements of all the applications. The Message Dispatcher concept was renamed as Message Handler (MH), and became an integral part of the SAE (Society of Automotive Engineers) J2735 DSRC Message Set Dictionary standard. It assumes that applications can require a different rate of BSMs, but does not consider the fact that each application could also require a different communication range. The solution defined in SAE J2735 focuses on adapting the rate of BSMs, but does not consider other application requirements (in particular the communication range) and does not exploit the flexibility offered by IEEE 802.11p to dynamically adapt other parameters such as the transmission power. By contrast, this paper proposes two techniques that exploit this flexibility to dynamically modify the transmission parameters in scenarios where vehicles simultaneously run multiple applications, and these applications have different requirements.

## III. PROBLEM DEFINITION

This study considers a scenario where each vehicle can simultaneously execute several applications. Each application has a set of requirements that need to be satisfied. It is then necessary to identify how many packets should be transmitted per second and their transmission power, so that the requirements of all the applications are efficiently and reliably satisfied. Different packets could be transmitted at different transmission power levels. This study considers as application requirements the communication range ($CR$) and the packet reception rate ($R$) or number of packets that need to be correctly received per second. These two requirements have been selected given their demonstrated relevance for safety applications [7] and their inclusion in cooperative ITS standards [5]. An application with requirements [$CR, R$] requires that at least $R$ packets are received per second at distances equal and lower than $CR$. Each vehicle runs simultaneously $N_A$ applications, and each application $j$ has its own communication range ($CR_j$) and packet reception rate ($R_j$) requirements ($1 \leq j \leq N_A$). These requirements can be changed with the context conditions. The objective of the proposed techniques is to dynamically configure the transmission parameters (packet rate or number of packets transmitted per second, and their transmission power level) in order to satisfy the application requirements and reduce the channel load. Lower channel load levels are important to avoid packet collisions and the saturation of the channel. Both effects have a very negative impact on the packet reception, and hence on the possibility to satisfy the application requirements [4].

The proposed techniques differ on how to achieve the described objectives. The first technique is referred to as MERLIN (optiMum powER and packet rate controL for multIple applicatioNs). MERLIN jointly considers the requirements of all the applications, and defines an optimization problem to find how many packets should be transmitted per second, and what their power level should be, in order to satisfy the requirements of all the applications and minimize the channel load. The problem is formulated as a constrained nonlinear optimization problem that is solved using the SQP (Sequential Quadratic Programming) method. MERLIN can produce optimum solutions at the expense of increasing the computational complexity and cost. The second technique is referred to as PRESTO (PoweR and mESsage raTe cOntrol for multiple applications). PRESTO has been designed to reduce the computational complexity and cost while trying to approximate the optimal solution. To this aim, PRESTO reduces the number of possible packet rates and transmission power levels, and searches for solutions to the optimization problem individually for each application. PRESTO then combines the solutions found for each application in order to identify how many packets should be transmitted per second, and what their transmission power should be to satisfy the requirements of all the applications.

## IV. MERLIN PROPOSAL

### a. Problem restrictions

We consider $N_A$ applications with requirements $[CR_j, R_j]$ for the communication range and the packet reception rate ($1 \leq j \leq N_A$). MERLIN is designed to calculate the vectors $\vec{T} = (T_1, T_2, ..., T_{N_V})$ and $\vec{P} = (P_1, P_2, ..., P_{N_V})$ that minimize the channel load generated while satisfying the applications requirements. $T_i$ denotes the number of packets per second that will be transmitted with the transmission power $P_i$ with $1 \leq i \leq N_V$, where $N_v$ is a parameter that represents the number of possible transmission power values. Each $T_i$ can take any value between 0 and $T_{max}$, and each $P_i$ can take any value between $P_{min}$ and $P_{max}$.

The application requirements represent restrictions to the optimization problem (one restriction per application). The solution of the optimization problem ($\vec{T}$ and $\vec{P}$) must satisfy the requirements of each application, i.e. the number of packets that are received at a distance $CR_j$ needs to be equal or higher than the packet reception rate $R_j$ demanded by application $j$, for $1 \leq j \leq N_A$. To obtain the problem restrictions, we need to model the number of packets that would be received at a given distance as a function of $\vec{T}$ and $\vec{P}$. To this aim, we consider that packets are independently received. The number of packets received every second follows a Binomial distribution $B(T, \rho)$, where $T$ is the number of packets transmitted per second, and $\rho$ represents the PDR (Packet Delivery Ratio), i.e. the probability of correctly receiving a packet. If a vehicle transmits $T$ packets per second with a transmission power $P$, the average number of packets that would be correctly received per second at a distance $d$ is:

$$\overline{r(d,P,C,T)} = T \cdot PDR(d,P,C) \quad (1)$$

where the PDR is a function of the distance $d$, the transmission power $P$ and the channel load experienced $C$. Different PDR

levels are experienced at different channel loads since the interference and packet collisions vary with the channel load. Satisfying on average the application requirements is not sufficient in vehicular networks. We propose then to use a lower bound of the number of packets that are correctly received per second. This bound is obtained using the Wilson score interval for binomial proportions [21]:

$$r(d,P,C,T) = \frac{T}{T+z^2}\left[\rho + \frac{z^2}{2T} - z\sqrt{\frac{\rho(1-\rho)}{T} + \frac{z^2}{4T^2}}\right] \quad (2)$$

where $z$ is the 1-α/2 percentile of a standard normal distribution, and $\rho = PDR(d,P,C)$ to simplify the notation. For a given $T$ and $\rho$, the number of packets correctly received every second will be lower than $r(d,P,C,T)$ with probability α/2.

Eq. (2) can be generalized to the case in which a vehicle transmits $T_i$ packets per second with a transmission power $P_i$ $\forall i$ with $1 \leq i \leq N_V$. In this case, the number of packets that a vehicle at distance $d$ to the transmitter would receive per second can be estimated as:

$$r(d,\vec{P},C,\vec{T}) = \sum_{i=1}^{N_V} r(d,P_i,C,T_i) \quad (3)$$

The application requirements are hence satisfied if the following conditions are met:

$$r(CR_j,\vec{P},C,\vec{T}) = \sum_{i=1}^{N_V} r(CR_j,P_i,C,T_i) \geq R_j \quad \forall j \quad 1 \leq j \leq N_A \quad (4)$$

Eq. (4) indicates that the lower bound of the number of packets correctly received at distance $CR_j$ must be higher or equal than the required packet reception rate $R_j$ $\forall j$ and $1 \leq j \leq N_A$. Eq. (4) represents the restrictions to the optimization problem.

### b. Objective function and optimization problem

From all possible solutions ($\vec{T}$ and $\vec{P}$) that satisfy the restrictions defined by eq. (4), MERLIN selects the one that minimizes the channel load generated by each vehicle. This load is here computed using the channel occupancy footprint (or footprint) introduced in [22]. This metric was proposed to compare the channel load generated by vehicles using different transmission parameters. The footprint is defined as the total channel resources consumed by the radio of a single vehicle in both time and space. It is calculated as the spatial integral of the channel load contribution of a vehicle [22]:

$$footprint = \int_{d=-\infty}^{d=+\infty} load(d,P,T) \quad (5)$$

As shown in (5), the load contribution of a vehicle depends on $P$ and $T$. As illustrated in Fig. 2, the same footprint can be generated when transmitting a low number of packets per second at a high transmission power, and when transmitting a higher number of packets per second but at a lower power. The load contributed at a distance $d$ by a vehicle transmitting $T$ packets per second at power $P$ can be computed as:

$$load(d,P,T) = T \cdot t_{pkt} \cdot PSR(d,P) \quad (6)$$

where $t_{pkt}$ represents the packet duration (in seconds) and $PSR(d,P)$ represents the Packet Sensing Ratio of a vehicle at distance $d$ from the transmitter that uses a transmission power $P$. The PSR is the probability of sensing (or detecting) a packet. Following [22], the PSR can be computed as the probability that a particular transmission is received by a node with a signal level higher than the carrier sense threshold, $CS_{Th}$. $CS_{Th}$ is the minimum received signal strength needed to detect a packet and therefore sense the channel as busy.

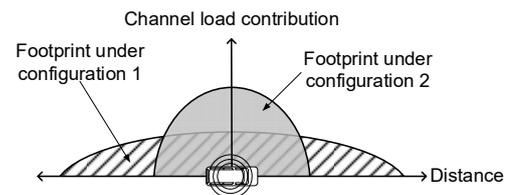

Fig. 2. Channel load and footprint of a vehicle with two configurations. Configuration 1: transmission of a low number of packets at high power. Configuration 2: transmission of a high number of packets at low power.

When a vehicle transmits $\vec{T}$ packets per second with a transmission power $\vec{P}$, the channel load contribution that it will generate at a distance $d$ can be expressed as:

$$load(d,\vec{P},\vec{T}) = t_{pkt} \cdot \sum_{i=1}^{N_V} T_i \cdot PSR(d,P_i) \quad (7)$$

Eq. (7) adapts the framework defined in [22] to multi-application scenarios, and considers the case in which vehicles can transmit packets with different power levels and packet rates. This is relevant when vehicles simultaneously run multiple applications, and each application has distinct requirements. It should be noted that we have developed our own models to calculate the footprint considering the WINNER+B1 propagation model. In particular, we have used our own PSR curves obtained by simulation. The footprint of a vehicle can be expressed as the spatial integral of the load it generates:

$$footprint = \int_{d=-\infty}^{d=+\infty} load(d,\vec{P},\vec{T}) \quad (8)$$

$$footprint = t_{pkt} \int_{d=-\infty}^{d=+\infty} \sum_{i=1}^{N_V} T_i \cdot PSR(d,P_i) = t_{pkt} \sum_{i=1}^{N_V} T_i \int_{d=-\infty}^{d=+\infty} PSR(d,P_i) \quad (9)$$

Finally, the problem can be formulated as a constrained nonlinear optimization problem for a given channel load experienced $C$ as:

$$\begin{aligned} \min \quad & footprint = t_{pkt} \sum_{i=1}^{N_V} T_i \int_{d=-\infty}^{d=+\infty} PSR(d,P_i) \\ s.t.: \quad & \sum_{i=1}^{N_V} r(CR_j,P_i,C,T_i) \geq R_j \quad \forall j \quad 1 \leq j \leq N_A \\ & 0 \leq T_i \leq T_{max} \quad 1 \leq i \leq N_V \quad T_i \in \mathbb{R} \\ & P_{min} \leq P_i \leq T_{max} \quad 1 \leq i \leq N_V \quad P_i \in \mathbb{R} \end{aligned} \quad (10)$$

The objective function and the problem restrictions are nonlinear equations. There are different methods to solve nonlinear constrained optimization problems. In this study, we have utilized the SQP method given its effectiveness and suitability for both small and large problems [23].

## V. PRESTO PROPOSAL

Solving the optimization problem identified in eq. (10) can result in computation times not acceptable for real-time systems, even when utilizing the SQP method. Vehicular networks require the ability to frequently and rapidly change the transmission parameters based on the application requirements and the context conditions. The proposal PRESTO has been designed with the objective to reduce the computation times while approximating the optimal solution. To this aim, PRESTO: 1) reduces the search space (i.e. the number of possible packet rates and transmission power levels); 2) identifies for each application individually the transmission parameters (packet rate and transmission power level) that satisfy the application requirements and minimize the channel load; and 3) combines the transmission parameters identified for each application to find a set of transmission parameters that satisfies the requirements of all applications.

PRESTO reduces the search space by configuring all the packets that are transmitted by a given application with the same transmission power level. In particular, PRESTO searches for each application for the optimum $\vec{T}$ and $\vec{P}$ considering that $N_V$ is equal to 1, i.e. vectors $\vec{T}$ and $\vec{P}$ are reduced to scalar variables $T$ and $P$. The objective is to minimize the channel load and satisfy the requirements of each application individually. PRESTO reduces further the search space by discretizing the possible values of $T$ and $P$. In particular, PRESTO reduces the search space for each application to a set of discrete $T$ values between 0 and $T_{max}$ with a transmission rate resolution of $\Delta T$, and a set of discrete $P$ values between $P_{min}$ and $P_{max}$ with a power resolution of $\Delta P$. The number of discrete $T$ values is equal to $N_T=1+T_{max}/\Delta T$, and the number of discrete $P$ values is equal to $N_P=1+(P_{max}-P_{min})/\Delta P$. Finding the optimum discretized variables $T$ and $P$ is equivalent to finding the optimum integer variables $k_T$ and $k_P$ in:

$$T = k_T \cdot \Delta T \quad 0 \leq k_T \leq (N_T-1) \quad k_T \in \mathbb{N} \quad (11)$$

$$P = P_{min} + k_P \cdot \Delta P \quad 0 \leq k_P \leq (N_P-1) \quad k_P \in \mathbb{N} \quad (12)$$

The optimization problem solved by PRESTO for each application $j$ is then expressed as follows for a given channel load experienced $C$:

$$\begin{aligned} \min \quad & footprint = t_{pkt} \cdot T \cdot \int_{d=-\infty}^{d=+\infty} PSR(d,P) \\ s.t.: \quad & r(CR_j,P,C,T) \geq R_j \\ & T = k_T \cdot \Delta T \quad 0 \leq k_T \leq (N_T-1) \quad k_T \in \mathbb{N} \\ & P = P_{min} + k_P \cdot \Delta P \quad 0 \leq k_P \leq (N_P-1) \quad k_P \in \mathbb{N} \end{aligned} \quad (13)$$

The pseudo-code of PRESTO is presented in Algorithm I, and is next explained. PRESTO considerably reduces the size of the search space (i.e. the number of possible combinations of packet rates and transmission power levels) compared to MERLIN. Such reduction is not only due to the discretization of the packet rate and transmission power levels, but also because PRESTO splits the optimization problem into

ALGORITHM I. PRESTO PROPOSAL

Inputs: $CR_j$ and $R_j$ $\forall j$ with $1 \leq j \leq N_A$

Output: vectors with optimum transmission parameters ($\vec{T}^*$ and $\vec{P}^*$)

1. **For** each application $j$ with $1 \leq j \leq N_A$ **do**
2.    Initialize $f_j$ to 1000
3.    **For** each $P$ in $P_{min}+\Delta P$, $P_{min}+2\cdot\Delta P$, …, $P_{min}+(N_P-1)\cdot\Delta P$ **do**
4.       **For** each $T$ in $\Delta T$, $2\cdot\Delta T$, …,$(N_T-1)\cdot\Delta T$ **do**
5.          Calculate $r(CR_j,P,C,T)$ with equation (2)
6.          **If** $r(CR_j,P,C,T) \geq R_j$ **then**
7.             Calculate footprint using equations (5-6)
8.             **If** the footprint is lower than $f_j$ **then**
9.                Set $f_j$ equal to the footprint calculated
10.               Save $P$ and $T$ as potential optimum for applic. $j$
11.             **End If**
12.          **End If**
13.       **End For**
14.    **End For**
15.    Set $P_j$ equal to the last $P$ saved in line 10
16.    Set $T_j$ equal to the last $T$ saved in line 10
17. **End For**
18. **Sort** $T_j$ and $P_j$ values in decreasing power levels
19.    Output: $\dot{P}_1 \geq \dot{P}_2 \geq … \geq \dot{P}_{N_A}$ and associated $\dot{T}_j$
20. Set $T^*_1$ equal to $\dot{T}_1$ and $P^*_1$ equal to $\dot{P}_1$
21. Set $t$ equal to $T^*_1$
22. **For** each application $j$ with $2 \leq j \leq N_A$ **do**
23.    Set $P^*_j$ equal to $\dot{P}_j$
24.    **If** $\dot{T}_j > t$ **then**
25.       Set $T^*_j$ equal to $\dot{T}_j - t$
26.       Set $t$ equal to $t + T^*_j$
27.    **Else**
28.       Set $T^*_j$ equal to 0
29.    **End If**
30. **End For**

multiple smaller sub-problems (one per application). The search space is equal to $N_P \cdot N_T$ for PRESTO, and equal to $N_P^{N_V} \cdot N_T^{N_V}$ for MERLIN if we consider discrete $T_i$ and $P_i$ values. For $N_P=50$ and $N_T=200$, the size of the search space is approximately equal to $10^{24}$ for MERLIN (with $N_V=6$), and equal to $N_P \cdot N_T=50 \cdot 200=10^4$ for PRESTO. In this context, it is possible for PRESTO to simply search through all possible combinations of $P$ and $T$ (i.e. all combinations of $k_T$ and $k_P$) to find a solution for each application that approximates the optimum one within a reasonable time (lines 1-17 of Algorithm I). For each potential solution ($P$, $T$), PRESTO calculates the number of packets that a given vehicle would receive per second at a distance $CR_j$ to the transmitter using Eq. (2) (line 5 of Algorithm I). If this number is higher or equal than the application requirement $R_j$, the solution is feasible (i.e. it satisfies the application requirements) and the footprint associated to this solution is computed using Eq. (5-6) (line 7 of Algorithm I). This footprint is compared to the minimum footprint computed for all the previous solutions. If it is smaller, then the solution under evaluation is selected as a possible optimum one (line 10 of Algorithm I). It will be the optimum solution if no other solution in the remaining search space can further reduce the footprint while satisfying the application requirements.

PRESTO follows the previous process to find the optimum configuration of transmission parameters for each application. It then combines the configurations of all the applications (lines 18-30 of Algorithm I) in order to minimize the number of packets to be transmitted while satisfying the requirements of all applications. The combination process first ranks the pairs of parameters $(P_j,T_j)$ of each application $j$ ($1 \leq j \leq N_A$) as a function of their transmission power (lines 18-19 of Algorithm I). The output is referred to as $(\dot{P}_j, \dot{T}_j)$, where $\dot{P}_1$ is the highest power and $\dot{T}_1$ is its associated packet rate. PRESTO then obtains the vector $\vec{P}^* = (P^*_1, P^*_2, ..., P^*_{N_A})$ of transmission power levels, and the vector $\vec{T}^* = (T^*_1, T^*_2, ..., T^*_{N_A})$ of packet rates (lines 20-30 of Algorithm I). Starting with $T^*_1=\dot{T}_1$ and $P^*_1=\dot{P}_1$, the number of packets transmitted per second using $\dot{P}_j$ (for $j>1$) is computed as the difference between $\dot{T}_j$ and the sum of the previously computed $T^*_j$ if such a difference is higher than zero, or zero otherwise. It is interesting to note that with this process the total number of packets that would be transmitted per second to satisfy all application requirements is $T=\max(T_1, T_2, …, T_{N_A})$. To illustrate the benefits of the proposed combination, let's consider an example with $N_A=2$ applications. Let's suppose that application 1 requires transmitting $\dot{T}_1=3$ packets per second at $\dot{P}_1=15$dBm, and application 2 requires transmitting $\dot{T}_2=5$ packets per second at $\dot{P}_2=10$dBm. Transmitting $\dot{T}_1+\dot{T}_2=8$ packets per second ($\dot{T}_1$ at $\dot{P}_1$ and $\dot{T}_2$ at $\dot{P}_2$) is unnecessary and inefficient since the requirements of both applications can be satisfied with only 3 packets transmitted per second at 15dBm and $\dot{T}_2-\dot{T}_1=2$ packets transmitted per second at 10dBm (i.e. with only 5 packets per second in total). This is exactly the result of PRESTO's combination process. PRESTO is therefore able to reduce the total number of packets transmitted and the channel load compared to a solution that does not perform any combination.

## VI. EVALUATION

### a. Scenario and settings

The proposed techniques have been evaluated using the network simulator ns2.35. The ns2 simulator models a 5 km straight road with 4 lanes (2 lanes per driving direction), and takes as input vehicular traces generated with the road traffic simulator SUMO (Simulation of Urban Mobility). Using SUMO, vehicles are periodically generated at both edges of the highway, and move with a maximum speed limit of 120km/h. Statistics are only collected for vehicles close to the center of the scenario in order to avoid boundary effects. Table 1 summarizes the main simulation parameters considered. Road traffic densities of 10, 20 and 30 vehicles/km/lane have been simulated. Each vehicle simultaneously executes $N_A$ applications. Each application belongs to an application class: class A includes applications that require a short communication range (0-80m) and high packet reception rate (7-10Hz); class B includes applications that require a medium communication range (80-160m) and medium packet reception rate (4-7Hz); class C includes applications with a larger communication range (160-240m) and low packet reception rate (1-4Hz). Each application is randomly assigned to a class at the beginning of the simulation and its specific communication range and packet reception rate are randomly selected within the application class ranges defined. We randomly select the application requirements so that our study is not constrained to specific applications and we can test our solutions for a large range of requirements. The proposed solutions could be adapted for dynamically adjusting the requirements to the vehicular context, e.g. the communication range could be adapted as a function of the vehicle speed. The radio propagation is modeled using the WINNER+ B1 model for urban environments with 10dB extra loss [24]. This model is recommended for V2V communications. The model has been implemented with an effective environment height of 1m to consider the effect of surrounding obstacles.

TABLE 1. SIMULATION PARAMETERS

| Parameter | Values |
| --- | --- |
| Number of applications ($N_A$) | 1, 2, 3, 4, 5 |
| α in eq. (2) | 0.05 |
| Data rate | 6 Mbps (QPSK ½) |
| Message size | 250 B |
| Carrier sense threshold of ($CS_{Th}$) | -85 dBm |
| Max / Min transmission power ($P_{max}$ / $P_{min}$) | 25 dBm / 0 dBm |
| Transmission power resolution ($\Delta P$) | 0.5dB |
| Number of possible powers ($N_P$) | 51 |
| Max / Min message rate ($T_{max}$ / $T_{min}$) | 20 Hz / 0 Hz |
| Message rate resolution ($\Delta T$) | 0.1 Hz |
| Number of possible message rates ($N_T$) | 201 |
| Size of vectors $\vec{T}$ and $\vec{P}$ in MERLIN ($N_V$) | $2 \cdot N_A$ |

The performance of MERLIN and PRESTO is compared to that obtained with the MH that is included in the SAE J2735 DSRC Message Set Dictionary standard. In multi-application scenarios, MH configures the packet rate as the minimum required to satisfy the packet rate requirements of all the applications, i.e. the packet rate is fixed to $T=\max(R_1, R_2, …, R_{NA})$. MH considers a fixed transmission power irrespective of the communication range requirements ($CR_i$). We fix in this study the MH transmission power to 25dBm.

MERLIN and PRESTO are executed at the beginning of the simulation to obtain the transmission parameters of each vehicle for the requirements randomly selected for the $N_A$ applications. The transmission power levels can vary between 0dBm and 25dBm (in steps of 0.5dB for PRESTO), and the packet rates between 0Hz and 20Hz (in steps of $\Delta T$=0.1Hz for PRESTO). MERLIN and PRESTO make use of models that relate the PDR and PSR with the distance, the transmission power and the channel load ($\rho=PDR(d,P,C)$ and $PSR(d,P)$ in Eq. (2), (10) and (13)). These models have been obtained with a spatial resolution of 10m for all the transmission power levels and 8 different channel load levels. Vehicles use different *PDR* curves depending on the channel load experienced. These models were obtained by simulating a wide range of scenarios with vehicles using different packet transmission rates to generate different channel load levels. It is important to highlight that these *PDR* curves include the packet loses due to propagation and packet collisions (i.e. due to interference and hidden terminal). Fig. 4 shows, as an example, the PDR curves for *P*=25dBm and Channel Busy Ratio (CBR) values from 0.1 to 0.8 in steps of 0.1. The CBR is used to measure the channel load experienced. It represents the percentage of time that a vehicle senses the channel as busy and is frequently used in congestion control studies in vehicular networks [4]. While the footprint is a metric that represents the channel load generated by a single vehicle, the CBR represents the channel load experienced by a vehicle as a result of radio transmissions from multiple vehicles.

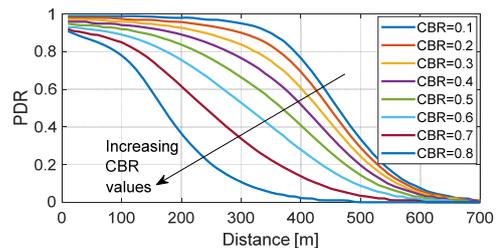

Fig. 4. PDR models for a transmission power of *P*=25dBm.

*b. Performance*

Fig. 5 compares the average CBR experienced when running each of the techniques under evaluation. The figure shows that PRESTO can reduce the CBR between 26% and 54% in multi-application scenarios compared to the standardized MH scheme. MERLIN further reduces the CBR, but the difference with PRESTO is not very significant. PRESTO is then able to closely approximate the CBR levels obtained with the optimum solution represented by MERLIN. The results depicted in Fig. 5 clearly demonstrate that the proposed techniques can significantly reduce the channel load experienced in multi-application scenarios compared to the standardized MH solution. This is very significant since one of the main challenges faced by vehicular networks is the channel congestion and the network stability.

Reducing the channel load should not be done at the expense of satisfying the application requirements. The capacity to satisfy the application requirements is evaluated in this study by means of the SAR (percentage of vehicles with Satisfied Application Requirements) metric. SAR is an adaptation to multi-application scenarios of the NAR (Neighborhood Awareness Ratio) metric proposed in [25]. NAR was defined in [25] as the proportion of vehicles within a specific range from which a packet is received during a time interval. The SAR metric is here defined as the percentage of vehicles from which the number of packets received during a time interval (set to one second in this study) is higher or equal than the number of packets required by the applications. The application requirements are satisfied for all vehicles if SAR is equal to 100%. Fig. 6 depicts the SAR metric obtained for the different techniques under evaluation. The figure shows that MERLIN and PRESTO significantly outperform MH. MERLIN and PRESTO are able to provide SAR levels above 99% for all traffic densities and number of applications run per vehicle. On the other hand, MH can only achieve SAR values between 70% and 95%. The degradation experienced with MH is due in part to the higher CBR levels depicted in Fig. 5. Higher CBR levels increase the interference, and hence the packet losses that degrade the capacity to satisfy the application requirements. Fig. 5 shows that the CBR significantly increases for MH with the traffic density. This explains the SAR degradation observed for MH when the traffic density increases (Fig. 6). Fig. 6 also shows that the SAR metric decreases for MH as the number of applications decreases. MH is configured to try to satisfy the application with the highest required packet reception rate. This configuration also satisfies the applications with lower requirements that actually might receive more packets than initially needed. If a vehicle runs only one application, the loss of a packet will degrade the SAR. However, if a vehicle runs more than one application, the loss of a packet results in that the most demanding application is not satisfied, but the less demanding applications could still be satisfied. This explains the higher SAR values observed for MH in Fig. 6 as the number of applications increases.

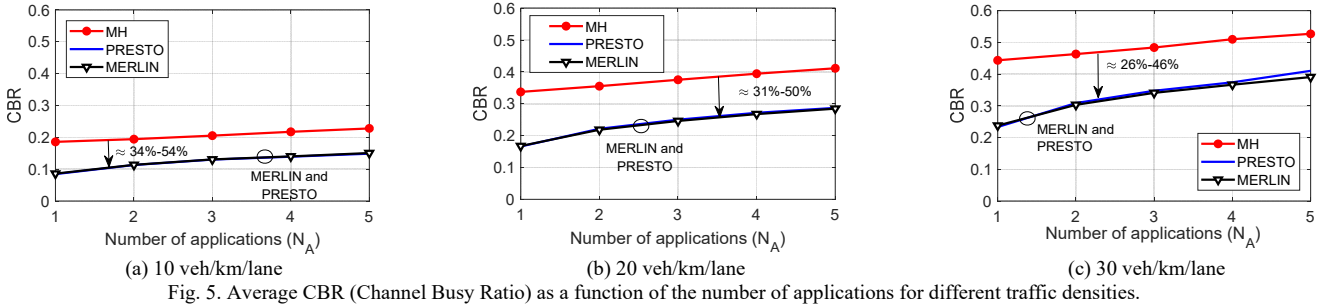

Fig. 5. Average CBR (Channel Busy Ratio) as a function of the number of applications for different traffic densities.

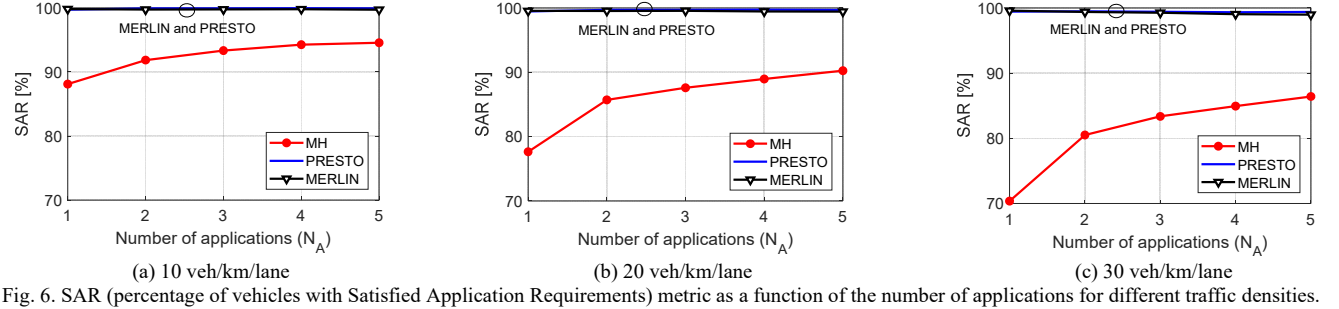

Fig. 6. SAR (percentage of vehicles with Satisfied Application Requirements) metric as a function of the number of applications for different traffic densities.

Fig. 5 and Fig. 6 clearly show that MERLIN and PRESTO significantly outperform MH in terms of channel load and capacity to satisfy the application requirements. These gains are a result of the capacity of PRESTO and MERLIN to dynamically adapt the transmission power and packet rate in order to satisfy the application requirements and reduce the channel load. On the other hand, MH uses a fixed transmission power, and sets the packet rate equal to the packet reception rate of the most demanding application. The operation of each technique is illustrated in Fig. 7 that plots the PDF (Probability Density Function) of the transmission power levels and packet rates used by all vehicles in the scenario when $N_A$=3. The x-axis in the packet rate PDFs represents the total number of packets transmitted per second per vehicle. Fig. 7 shows that MERLIN and PRESTO significantly reduce the transmission power compared to MH, but increase the number of packets transmitted per second. Compared to MH, MERLIN and PRESTO can therefore benefit applications that need to communicate with nearby vehicles, and that require a high packet rate given the proximity of potentially colliding vehicles. In addition, reducing the transmission power can reduce unnecessary interferences created to vehicles at very large distances.

Fig. 8 plots the DP (Packets Difference) metric as a function

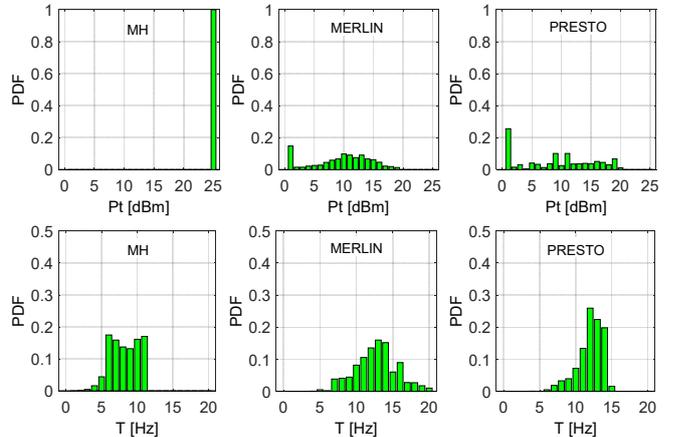

Fig. 7. PDF (Probability Density Function) of transmission power and packet rate used by vehicles in the scenario when $N_A$=3.

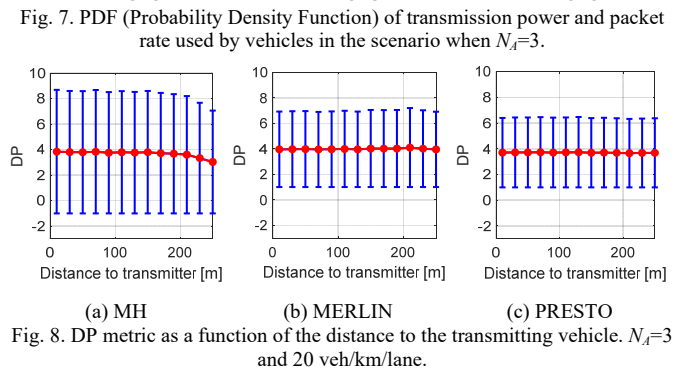

Fig. 8. DP metric as a function of the distance to the transmitting vehicle. $N_A$=3 and 20 veh/km/lane.

      

of the distance to the transmitting vehicle when $N_A$=3 and the traffic density is 20 veh/km/lane. The DP metric represents the difference between the number of packets correctly received per second, and the number of packets per second required by the applications. The figure plots the average values and the 5[th] and 95[th] percentiles that are represented with vertical lines. The application requirements are not satisfied if the DP metric is lower than zero. The figure shows that MH is not capable to satisfy the application requirements of all vehicles irrespective of the distance between transmitter and receiver. MERLIN and PRESTO achieve 5[th] percentile values higher than zero for all distances. Vehicles are always satisfied since they receive more packets per second than required by the application. Compared to MH, MERLIN and PRESTO better adjust the number of packets received per second and the transmission power to the application requirements for all vehicles.

### c. Computational cost and complexity

MERLIN and PRESTO significantly reduce the channel load and improve the capacity to satisfy the application requirements compared to MH. Fig. 5 and Fig. 6 showed that PRESTO can achieve similar performance levels to MERLIN. In addition, it can reduce the complexity and computational cost. The complexity of solving the problem formulated in MERLIN with the SQP method is of order $O(N_A^3)$ [26]. PRESTO reduces the complexity[1] to $O(N_A^2)$.

Table 2 shows the number of CPU cycles needed to execute each line of the PRESTO algorithm. The *Repetitions* column represents how many times each line needs to be executed (many are inside *for* loops). It is actually an upper bound since some lines are only executed when a certain condition is satisfied (e.g. lines 7-10). The number of CPU cycles has been computed considering Intel CPU architectures [27]. For example, the multiplication of two floating point numbers requires 5 CPU cycles, and their division requires 39 cycles. In the worst case, the sorting process represented in lines 18-19 has $N_A \cdot N_A$ iterations [28] since $N_A$ is the size of the vector to be sorted. We assume that each iteration requires 3 CPU cycles (one comparison and two substitutions). Using Table 2 we can estimate an upper bound of the total number of CPU cycles needed to run PRESTO. This number is approximately equal to $8 \cdot 10^6$ CPU cycles[2] when we consider $N_A$=3 applications and the values of $N_P$ and $N_T$ reported in Table 1.

Fig. 9 plots the upper bound of the CPU time needed to run PRESTO for different CPU speeds and a varying number of applications. The MK5-OBU unit of Cohda Wireless currently has an embedded 1GHz ARM Cortex-A9 processor [29]. Audi has announced that its vehicles will integrate the Qualcomm® Snapdragon™ 602A processors with a CPU of 1.5GHz [30]. Volkswagen will integrate in its 2019 vehicles the latest Qualcomm® Snapdragon™ 820A automotive processor with a CPU of up to 2.2GHz [31]. Fig. 9 shows that in the worst-case scenario (1GHz CPU and $N_A$=5), the upper bound of the CPU time needed to execute PRESTO is approximately 14ms. These results demonstrate that it is feasible to implement PRESTO in real systems, in particular if we take into account that PRESTO only needs to be executed when the application requirements change. It is also important to note that the CPU time could be decreased by simply reducing the number of possible transmission power levels ($N_P$) or packet rates ($N_T$). For example, halving the possible values of $N_P$ or $N_T$ would approximately halve the required CPU time. However, this is achieved at the expense of reducing the precision with which PRESTO approximates the CBR of the optimum solution obtained with MERLIN.

TABLE 2. COMPUTATIONAL COST OF THE PRESTO TECHNIQUE

| Algorithm line | CPU cycles | Repetitions |
|---|---|---|
| 1 | 2 | $N_A$ |
| 2 | 1 | $N_A$ |
| 3 | 4 | $N_A \cdot N_P$ |
| 4 | 4 | $N_A \cdot N_P \cdot N_T$ |
| 5 | 257 | $N_A \cdot N_P \cdot N_T$ |
| 6 | 1 | $N_A \cdot N_P \cdot N_T$ |
| 7 | 6 | $N_A \cdot N_P \cdot N_T$ |
| 8 | 1 | $N_A \cdot N_P \cdot N_T$ |
| 9 | 1 | $N_A \cdot N_P \cdot N_T$ |
| 10 | 2 | $N_A \cdot N_P \cdot N_T$ |
| 15 | 1 | $N_A$ |
| 16 | 1 | $N_A$ |
| 18-19 (sorting) | 3 | $N_A \cdot N_A$ |
| 20 | 2 | 1 |
| 21 | 1 | 1 |
| 22 | 2 | $N_A$-1 |
| 23 | 1 | $N_A$-1 |
| 24 | 1 | $N_A$-1 |
| 25 | 5 | $N_A$-1 |
| 26 | 2 | $N_A$-1 |

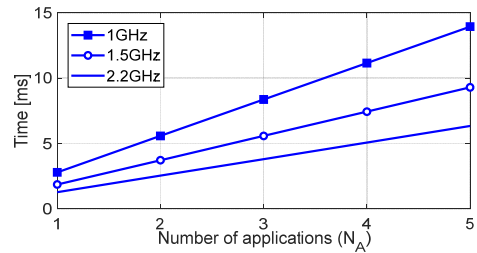

Fig. 9. Upper bound of the CPU time needed to execute PRESTO with the values of $N_P$ and $N_T$ reported in Table 1.

As an example, Fig. 10 plots the CDF (Cumulative Distribution Function) of the time needed to run MERLIN and PRESTO using an Intel Xeon CPU of 1.8GHz for $N_A$=3. While MERLIN required between 3s and 11s to find the optimum solution, the time needed to run PRESTO was between 1ms and 3ms. PRESTO is then 3 orders of magnitude faster than MERLIN. The results depicted in Fig. 9 and Fig. 10 confirm that it is possible to execute PRESTO in real-time using current CPUs.

---

[1] The complexity depends on the process used to sort the power and packet rates found (lines 18-19 of Algorithm I) which has a complexity of order $O(N_A^2)$. A different sorting process could reduce this complexity.

[2] We assume that the spatial integral of the PSR is pre-computed for each transmission power level.

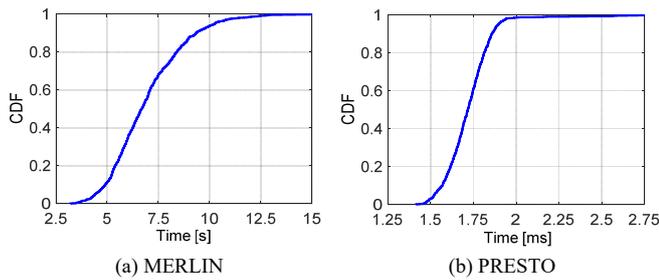

(a) MERLIN  (b) PRESTO

Fig. 10. CDF of the execution time required to run MERLIN and PRESTO for $N_A$=3. Results obtained using an Intel Xeon CPU of 1.8GHz.

## VII. CONCLUSIONS

Connected vehicles will simultaneously run multiple applications that can have different requirements. The communication parameters should be carefully configured in order to guarantee that all application requirements are satisfied while controlling the channel load to ensure the networks' scalability. In this context, this paper has proposed two novel techniques designed to dynamically configure the communication parameters of vehicles simultaneously executing multiple applications. To the authors' knowledge, the proposed techniques are the first to adapt the transmission power and packet rate in multi-application scenarios. The two techniques have been designed with the objective to satisfy all the application requirements and reduce the channel load. They differ on how they tradeoff between an optimum configuration of communication parameters and the computational cost. The obtained results demonstrate that the proposed techniques significantly improve the performance achieved with the MH included in the SAE J2735 DSRC Message Set Dictionary standard.

## REFERENCES

[1] Z. MacHardy et al., "V2X Access Technologies: Regulation, Research, and Remaining Challenges", *IEEE Communications Surveys & Tutorials*, vol. 20, no. 3, pp. 1858-1877, third quarter 2018.

[2] IEEE 802.11p-2010: Wireless LAN MAC and PHY Specifications Amendment 6: Wireless Access in Vehicular Environments, *IEEE Standards Association*, 2010.

[3] ETSI TC ITS, "Intelligent Transport Systems (ITS); European Profile Standard for the Physical and Medium Access Control Layer of Intelligent Transport Systems operating in the 5 GHz frequency band", ES 202 663, 2010.

[4] M. Sepulcre, J. Gozalvez, Onur Altintas and Haris Kremo, "Integration of Congestion And Awareness Control in Vehicular Networks", *Ad Hoc Networks*, vol. 37, part 1, pp. 29–43, Feb. 2016.

[5] ETSI TC ITS, "Intelligent Transport Systems (ITS); Vehicular Communications; Basic Set of Applications; Definitions", *ETSI TR 102 638 V1.1.5*, Jan. 2016.

[6] ETSI TC ITS, "Intelligent Transport Systems (ITS); Vehicular Communication, Basic Set of Applications, Part 4: Operational Requirements", *Draft ETSI TS 102 637-4*, March 2010.

[7] M. Sepulcre and J. Gozalvez, "On the importance of application requirements in cooperative vehicular communications", *Proc. Int. Conf. on Wireless On-Demand Network Systems and Services (WONS)*, pp.124-131, Bardonecchia, Italy, 26-28 Jan. 2011.

[8] H. S.Basheer, C. Bassil, "A review of broadcasting safety data in V2V: Weaknesses and requirements", *Ad Hoc Networks*, vol. 65, pp. 13-25, October 2017.

[9] B. M. Masini, A. Bazzi and A. Zanella "A Survey on the Roadmap to Mandate on Board Connectivity and Enable V2V-Based Vehicular Sensor Networks", *Sensors*, 18(7), 2207, 2018.

[10] M. Sepulcre et al., "Contextual Communications Congestion Control for Cooperative Vehicular Networks", *IEEE Transactions on Wireless Communications*, vol. 10 no. 2, Feb. 2011.

[11] P. Alexander et al., "Cooperative Intelligent Transport Systems: 5.9-GHz Field Trials", *Proceedings of the IEEE*, vol. 99, no. 7, pp. 1213-1235, July 2011.

[12] D. Tian et al., "Self-Organized Relay Selection for Cooperative Transmission in Vehicular Ad-Hoc Networks," *IEEE Transactions on Vehicular Technology*, vol. 66, no. 10, pp. 9534-9549, Oct. 2017.

[13] M. Sepulcre et al., "Cooperative vehicle-to-vehicle active safety testing under challenging conditions", *Transportation Research Part C: Emerging Technologies*, vol. 26, pp. 233-255, Jan. 2013.

[14] C.L. Huang et al., "Intervehicle Transmission Rate Control for Cooperative Active Safety System", *IEEE Transactions on Intelligent Transportation Systems*, vol. 12, no. 3, pp. 645-658, Sept. 2011.

[15] M. Sepulcre, J.Gozalvez, "Coordination of Congestion and Awareness Control in Vehicular Networks", *Electronics* 7(11), 335, 2018.

[16] J. Gozalvez and M. Sepulcre, "Opportunistic technique for efficient wireless vehicular communications", *IEEE Vehicular Technology Magazine*, vol. 2, no. 4, pp. 33-39, Dec. 2007.

[17] S. Joerer et al., "Enabling Situation Awareness at Intersections for IVC Congestion Control Mechanisms", *IEEE Transactions on Mobile Computing*, vol. 15, no. 7, July 2016.

[18] Z. Y. Rawashdeh et al., "A Scalable Application and System Level-Based Communication Scheme for V2V Communications", *Proc. IEEE 84th Vehicular Technology Conference (VTC-Fall)*, Montreal (Canada), pp. 1-5, 18–21 Sept. 2016.

[19] L. Zhang, S. Valaee, "Congestion Control for Vehicular Networks With Safety-Awareness", *IEEE/ACM Transactions on Networking*, vol. 24, no. 6, pp. 3290-3299, Dec. 2016.

[20] C.L. Robinson et al., "Efficient Message Composition and Coding for Cooperative Vehicular Safety Applications", *IEEE Transactions on Vehicular Technology*, vol. 56, no. 6, pp. 3244-3255, Nov. 2007.

[21] E. B. Wilson, "Probable inference, the law of succession, and statistical inference", *Journal of the American Statistical Association*. vol. 22, pp. 209–212, 1927.

[22] Qi Chen et al., "Mathematical Modeling of Channel Load in Vehicle Safety Communications", *IEEE Vehicular Technology Conference VTC-Fall*, San Francisco, USA, pp.1-5, 5-8 Sept. 2011.

[23] J. Nocedal and S. J. Wright, "Numerical Optimization", *Springer* (Second Edition), 2006.

[24] METIS Consortium, "Initial channel models based on measurements", *ICT-317669-METIS/D1.2*, April 2014.

[25] M. Boban et al., "Exploring the Practical Limits of Cooperative Awareness in Vehicular Communications", *IEEE Transactions on Vehicular Technology*, vol. 65, no. 6, pp. 3904-3916, June 2016.

[26] R. Ghaemi, "Robust Model Based Control of Constrained Systems", PhD Dissertation, University of Michigan 2010.

[27] Intel, "Intel® 64 and IA-32 Architectures Optimization Reference Manual", Order Number: 248966-033, June 2016.

[28] T. H. Cormen et al., "Introduction to Algorithms", MIT Press, 2009.

[29] Cohda Wireless MK5 V2X On Board Unit. Online: http://cohdawireless.com [Last access May 2019]

[30] Qualcomm Press: "Audi 2017 Vehicles to Integrate Qualcomm Snapdragon 602A Infotainment Processor", May 2019. Online: https://www.qualcomm.com/news/releases/2016/01/05/audi-2017-vehicles-integrate-qualcomm-snapdragon-602a-infotainment

[31] Qualcomm Press: "Qualcomm Automotive Solutions Power Next Generation Infotainment for Volkswagen Vehicles", May 2019: https://www.qualcomm.com/news/releases/2017/01/03/qualcomm-cutting-edge-automotive-solutions-power-next-generation